\documentstyle[12pt]{article}
\renewcommand{\title}[1]{    \bigskip
    \begin{center}    \Large\bf #1    \end{center}    \vskip .2cm}

\renewcommand{\author}[1]{%
    {\begin{center}
    #1
    \end{center}}}
\newcommand{\address}[1]{\vspace{-1.7em}\vspace{0pt}
    {\begin{center}
    \it #1
    \end{center}}}
\begin{document}


\title{On the question of regular solitons in a Noncommutative Maxwell--Chern--Simons--Higgs model}

\author{
Pradip Mukherjee\footnote{Also Visiting Associate, S. N. Bose National Centre 
for Basic Sciences, JD Block, Sector III, Salt Lake City, Calcutta -700 098, India and\\ IUCAA, Post Bag 4, Pune University Campus, Ganeshkhind, Pune 411 007,India }
$\,^{\rm a,b}$\, Anirban Saha $\,^{\rm a,c}$ }
\address{$^{\rm a}$Department of Physics, Presidency College\\
86/1 College Street, Kolkata - 700 073, India}
\address{$^{\rm b}$\tt pradip@bose.res.in}
\address{$^{\rm c}$\tt ani\_saha09@dataone.in}

\begin{abstract}The Maxwell--Chern--Simons model with scaler matter in the adjoint representation is analyzed from an alternative approach which is regular in the $\theta \to 0$ limit. This method is complementary to the usual operator formalism applied to explore the nonperturbative solutions which gives singular results in the $\theta \to 0$ limit. The absence of any regular non-trivial lumpy solutions satisfying B--P--S bound has been conclusively demonstrated.

\end{abstract}

\noindent {\bf PAC codes:} 11.15.-q, 11.10.Nx,  \\
{\bf{Keywords:}} Maxwell--Chern--Simons--Higgs Model, Solitons, Noncommutativity


\vspace{1.0cm} 
Gauge theories defined over noncommutative (NC) space, where the coordinates $x^{\mu}$ satisfy the Heisenberg algebra $\left\{x^{\mu},x^{\nu}\right\} = i \theta^{\mu \nu}$ with $\theta^{\mu \nu}$ a constant antisymmetric tensor\footnote{Heisenberg first suggested this idea which was later developed by Snyder \cite{sny1, sny2}.}, has become prominent in the recent literature. The principal motivation came from the emergence of such field theories from the context of string theory \cite{SW}. Field theories formulated on NC platform have later been established in their own right \cite{szabo}. The essential fuzziness of space time associated with noncommutativity has led to many nontrivial solutions of theories which are otherwise trivial in the commutative limit. New interactions originate \cite{sch}, gauge couplings develop \cite{ms} and various other phenomena emerge carrying NC signature. A particularly interesting scenario in this context is the occurence of classical stable soliton solutions in odd dimensional scalar NC field theories with self interaction only \cite{gms} and also in a NC $U(1)$ gauge theory coupled with matter in the adjoint representation of the gauge group \cite{poly}. The NC $U(1)$ gauge theories coupled with adjoint matter is important for their possible application in constructing D-branes as solitons of the tachyon field in NC open string theory \cite{hkl,as}. It is indeed worthwhile to analyse such theories from different points of view.  

 There are basically two broad approaches of treating NC field theories. One approach is to work in the deformed phase space where the ordinary product of two functions ${\hat{\phi}}(x)$ and ${\hat{\psi}}(x)$ is replaced by their star product
\begin{equation}
\hat \phi(x) \star \hat \psi(x) = \left(\hat \phi \star \hat \psi \right)(x) = e^{\frac{i}{2}\theta^{\alpha\beta}\partial_{\alpha}\partial^{'}_{\beta}}
\hat \phi (x) \hat \psi(x^{'})\big{|}_{x^{'}=x.}
\label{star}
\end{equation}
Alternatively, one works in a certain Hilbert space that carries a representation of the basic noncommutative algebra. The fields are promoted to operators in this Hilbert space, the form of which is usually determined by the Weyl--Wigner correspondence. For the case of adjoint matter coupled with $U(1)_{\star}$ Chern--Simons (CS) gauge field these different approaches have been shown to reveal mutually independent sectors of solutions of the theory \cite{ms}. In the present paper we will address the coupling of adjoint matter with the Maxwell field in addition to the C--S term. The problem was earlier addreseed in \cite{kp} from the operator approach. 

   A solution generating technique  based on the operator approach has been used in \cite{kp} which was developed in
\cite{hkl}. Here different solutions have been obtained in various limits of the action, 
\begin{equation}
S = \int dt d^{2}x \left( -\frac{1}{4} \left(F_{\mu \nu}\right)^{2} + \frac{1}{2}D^{\mu}\phi D_{\mu}\phi - V\left(\phi - \phi_{\star}\right)\right)
\label{model}
\end {equation}
where $\phi$ is the scalar field in the adjoint representation of the NC U(1) group. The potential $V$ has a local minima at $\phi = \phi_{\star}$ with $V(0)= 0$ and a local maximum at $\phi = 0$. The static soliton solution of the theory has energy
\begin{equation}
E = 2 \pi \theta n \left(\frac{1}{2 \theta^{2}} + V\left( - \phi_{\star}\right)\right)
\end {equation} 
which diverges as $\theta \to 0$. These soliton solutions obtained by using the operator approach thus essentially belong to the singular sector. It is not clear what happens in the $\theta \to 0$ limit.
Specifically, the question is whether there is some non-trivial non-perturbative solutions depending on the NC parameter and vanishing continuously along with it. Clearly, this question can not be answered by the operator approach. It is however very much desirable to explore all sectors of solutions of the theory. At this point our method of analysis \cite{ms} which is regular in the $\theta \to 0$ limit becomes instrumental. Note that this is also an exact method valid non-perturbatively which compares favourably with the operator approach.
 
 In the following we will consider the model 
\begin{equation}
\hat S = \int d^{3}x\left[-\frac{1}{4 g^{2}}\hat F_{\mu\nu}\star \hat F^{\mu\nu} + \frac{1}{2} \left(\hat D_{\mu} \star \hat \phi \right)\star \left(\hat D^{\mu} \star \hat \phi\right) + \frac{k}{2}\epsilon^{\mu \nu \lambda}\left(\hat A_{\mu} \star \partial_{\nu}\hat A_{\lambda} - \frac{2i}{3}\hat A_{\mu} \star \hat A_{\nu} \star \hat A_{\lambda}\right)\right]
\label{ncaction}
\end{equation}
 where $\hat \phi$ is the scalar field and $\hat A_{\mu}$ is the NC gauge field. We adopt the Minkowski metric $\eta_{\mu \nu} = {\rm diag} \left( +,-,-,-\right)$. The dynamics of the gauge field is assumed to be governed by a combination of the Maxwell and the C--S term and thus more general than (\ref{model}). 
Note that we have not included any potential term in (\ref{ncaction}). Our motivation is to find whether any non-trivial coupling is possible in the smooth $\theta \to 0$ sector. In the event of such coupling an appropriate potential term can be devised to saturate the BPS limits.
 
 The commutative equivalent of (\ref{ncaction}) can be easily obtained by using the exact S--W maps for $ D_{\mu}\star\hat \phi(x)$ and $-\frac{1}{4 g^{2}}\hat F_{\mu\nu}\star \hat F^{\mu\nu}$ given in \cite{hsk, rbhsk} and noting that the C--S action retains its form under SW map \cite{gs}. Proceeding in this direction we write the commutative equivalent of (\ref{ncaction}) as
\begin{eqnarray}
\hat S \stackrel{\rm{SW \; map}}{=}  \int d^{3}x \left[\sqrt{{\mathrm {det}} \left( 1 + F \theta \right)}\left\{\frac{1}{4 g^{2}}\left(\frac{1}{1+ F \theta}\right)^{\mu \alpha} F_{\alpha\beta}\left(\frac{1}{1+ F \theta}\right)^{\beta\nu} F_{\nu\mu}\right.\right.\nonumber\\
\qquad \qquad \qquad \qquad + \left.\left. \frac{1}{2}\left(\frac{1}{1 + F \theta }\frac{1}{1 + \theta F}\right)^{\mu \nu} \partial_{\mu} \phi\partial_{\nu} \phi\right\} + \frac{k}{2}\epsilon^{\mu \nu \lambda} A_{\mu} \partial_{\nu} A_{\lambda}\right]               
\label{caction}                
\end{eqnarray}
In (\ref{caction}) we have used the matrix notation
\begin{eqnarray}                                                          
\left(AB\right)^{\mu \nu} &=& A^{\mu \lambda}B_{\lambda}{}^{\nu} \nonumber\\       AB &=& A^{\mu \lambda}B_{\lambda}{}_{\mu}                                   
\label{matnot}
\end{eqnarray}
Also $\left(1 + F\theta \right)$ is to be interpreted as a mixed tensor in calculating the determinant. Note that the invariance of the theory under $U(1)$ gauge transformation is manifest in (\ref{caction}). 
It is now straightforward to write down the equations of motion for  the scalar field $\phi$ and the gauge field $A_{\mu}$ from (\ref{caction}) respectively as
\begin{equation}
\partial_{\alpha}\left\{\sqrt{{\mathrm {det}} \left( 1 + F \theta \right)}\left(\frac{1}{1 + F \theta }\frac{1}{1 + \theta F}\right)^{\alpha \nu} \partial_{\nu} \phi\right\} = 0
\label{eqmphi}
\end{equation}
and
\begin{eqnarray}
k\epsilon^{\alpha \nu \lambda} \partial_{\nu} A_{\lambda} &-& \frac{1}{4g^{2}}\partial_{\xi}\left[ \sqrt{{\mathrm {det}} \left( 1 + F \theta \right)}\left\{\frac{1}{2}\left(\theta\frac{1}{1+ F\theta}+\frac{1}{1+ \theta F}\theta\right)^{\alpha\xi}\left(\frac{1}{1+ F\theta} F \frac{1}{1+ F\theta} F\right) \right.\right.\nonumber \\
&+& \left.\left.2\left(\frac{1}{1+ F\theta} F \frac{1}{1+ F\theta}+ \frac{1}{1+ \theta F} F \frac{1}{1+ \theta F} \right)^{\alpha\xi}\right.\right.\nonumber \\ &-&\left.\left. 2\left(\theta \frac{1}{1+ F\theta} F \frac{1}{1+ F\theta} F \frac{1}{1+ F\theta}+ \frac{1}{1+ \theta F} F \frac{1}{1+ \theta F} F \frac{1}{1+ \theta F}\theta \right)^{\alpha\xi} \right\}\right] = j^{\alpha} \nonumber\\
\label{eqmA}
\end{eqnarray} 
where, 
\begin{eqnarray}
 j^{\alpha} &=& \partial_{\xi}\left[\sqrt{{\mathrm {det}} \left( 1 + F \theta \right)} \left\{\frac{1}{4} \left(\theta \frac{1}{1 + F \theta } + \frac{1}{1 + \theta F} \theta\right)^{\alpha \xi}\left(\frac{1}{1 + F \theta }\frac{1}{1 +\theta F}\right)^{\mu \nu} \right.\right.\nonumber\\
&& +  \left(\frac{1}{1 + F \theta }\frac{1}{1 + \theta F} \theta \right)^{\mu \alpha} \left(\frac{1}{1 + \theta F}\right)^{        \xi \nu} \nonumber\\
 &&+ \left. \left. \left(\frac{1}{1 + F \theta}\right)^{\mu \alpha} \left(\theta \frac{1}{1 + F \theta }\frac{1}{1 + \theta F} \right)^{\xi \nu} \right\} \partial_{\mu} \phi\partial_{\nu} \phi\right]
\label{defJ}
\end{eqnarray}
By direct computation from (\ref{defJ}) we get 
\begin{eqnarray}
\partial_{\alpha} j^{\alpha} = 0
\label{continuity}
\end{eqnarray}
So $j^{\alpha}$ in (\ref{defJ}) can be consistently interpreted as the matter current. If we go to the limit $g \to \infty$, the Maxwell term is eliminated from the action. In this limit we can compare the equations of motion with the results given in \cite{ms}. There, we have discussed elaborately, the issues of commutative limit and first order approximations. The observations in this context apply equally for the present model. 

 The equations (\ref{eqmphi}) and (\ref{eqmA}) are a set of coupled nonlinear equations. We will now investigate whether they admit any solitary wave solution. This can be seen in a systematic way by looking for the Bogomolnyi bounds of the equations. To this end we turn to the construction of the energy functional for which we require the physical energy momentum (EM) tensor corresponding to the model (\ref{caction}). The issue of energy-momentum tensor for a NC gauge theory involves many subtle points as evidenced in the literature \cite{bly}. We have shown in \cite{ms} that the canonical (Noether) procedure or the extended method of Jackiw \cite{jackiw1, jackiw2} do not give both symmetric and gauge invariant EM tensor. The best way is to find a symmetric and gauge-invariant EM tensor by varying the action (\ref{caction}) with respect to a background metric and finally keeping the metric flat. 
 
 We thus extend the action (\ref{caction}) as
\begin{eqnarray}
S = \int d^{3}x \sqrt{- g} \mathcal{L}
\label{lag}
\end{eqnarray}
 where $g = \mathrm{det} g_{\mu \nu}$ and  $g_{\mu \nu}$ is the background metric. By definition \cite{nallu} the symmetric gauge-invariant EM tensor is
\begin{eqnarray}\Theta^{\left(s\right)}_{\alpha\beta} = 2 \frac{\partial \mathcal{L}}{\partial g^{\alpha \beta}} - {\mathcal{L}} g_{\alpha \beta}
\label{SEMT}
\end{eqnarray}
in the limit $g_{\mu \nu} \to \eta_{\mu \nu}$. Explicitly
\begin{eqnarray}
\Theta^{\left(s\right)}_{\rho\lambda} &=& \sqrt{{\mathrm {det}}\left( 1 + F \theta \right)}\left[\frac{1}{4g^{2}}\left(\frac{1}{1 + \theta F}F\theta\right)_{\rho\lambda}\left(\frac{1}{1 + \theta F} F \frac{1}{1 + \theta F}F\right) 
+ \frac{1}{g^{2}}\left(\frac{1}{1 + \theta F} F \frac{1}{1 + \theta F}F\right)_{\rho\lambda} \right.\nonumber\\
&& - \left.\frac{1}{4g^{2}}\left(\frac{1}{1 + \theta F} F \frac{1}{1 + \theta F}F\right)g_{\lambda\rho} + \frac{1}{2}\left(\frac{1}{1 + \theta F}F\theta\right)_{\rho\lambda}\left(\frac{1}{1 + F \theta }\frac{1}{1 + \theta F}\right)^{\mu\nu}\partial_{\mu}\phi \partial_{\nu}\phi \right.\nonumber\\
&& \left.+ \left(\frac{1}{1 + F \theta }\frac{1}{1 + \theta F}\right)_{\rho}{}^{\nu}\partial_{\lambda}\phi \partial_{\nu}\phi - \frac{1}{2}\left(\frac{1}{1 + F \theta }\frac{1}{1 + \theta F}\right)^{\mu\nu}\partial_{\mu}\phi \partial_{\nu}\phi g_{\rho\lambda}
\right]
\label{symemt}
\end{eqnarray}
Naturally this EM tensor is physically significant and we can work out the energy density of the field system from the time-time component of this EM tensor.
Until now our approach was completely general in that we do not assume vanishing time-space noncommutativity. 
However, assuming $\theta^{0i} = 0$ is almost conventional in the study of NC solitons. In particular the calculations of non-perturbative solutions in NC $U(1)$ gauge theory reported in the literature assume noncommutativity only in the spatial directions.  Going over to this limit, the energy functional becomes
\begin{eqnarray}
E =  \int d^{3}x \Theta^{\left(s\right)}_{00} &=& \int d^{3}x \frac{1}{2} \sqrt{{\mathrm{det}} \left( 1 + F \theta \right)}\left\{-\frac{1}{2g^{2}}\left(\frac{1}{1 + F \theta } F \frac{1}{1 + F \theta }F\right) \right.\nonumber\\
&&+\left. \frac{2}{g^{2}}\left(\frac{1}{1 + F \theta } F \frac{1}{1 + F \theta }F\right)_{00}
+2\left(\frac{1}{1 + F \theta }\frac{1}{1 + \theta F}\right)_{0}{}^{\nu} \left( \partial_{0} \phi \partial_{\nu} \phi\right) \right.\nonumber\\
&&-\left.\left(\frac{1}{1 + F \theta }\frac{1}{1 + \theta F}\right)^{\mu \nu} \left( \partial_{\mu} \phi \partial_{\nu} \phi\right)\right\}
\label{statenrgy}
\end{eqnarray}
With the stated assumptions about NC tensor $\theta^{\mu \nu}$ the form of the matrices appearing in the above equation can be easily worked out. Explicitly the matrix for the $\left(\frac{1}{1 + F \theta }\frac{1}{1 + \theta F}\right)^{\mu \nu}$ can be written as 
\begin{eqnarray}
\left(\frac{1}{1 + F \theta }\frac{1}{1 + \theta F}\right)^{\mu \nu}
= \left(\begin{array}{ccc}
\left\{1 - \frac{\theta^{2}\left(E_{1}^{2} + E_{2}^{2}\right)}{\left(1 - \theta B\right)^{2}}\right\} & \frac{\theta E_{2}}{\left(1 - \theta B\right)^{2}}  & \frac{-\theta E_{1}}{\left(1 - \theta B\right)^{2}}  \\
\frac{\theta E_{2}}{\left(1 - \theta B\right)^{2}}  & \frac{-1}{\left(1 - \theta B\right)^{2}}  & 0 \\
\frac{-\theta E_{1}}{\left(1 - \theta B\right)^{2}} & 0 & \frac{-1}{\left(1 - \theta B\right)^{2}}\end{array}\right)
\label{thematrix}
\end{eqnarray}
Also ${\rm det}\left(1 + F\theta \right) = \left(1 - \theta B\right)^{2}$. 
In the static limit the energy functional becomes
\begin{equation}
E = \int d^{2}x \frac{\sqrt{{\mathrm{det}} \left( 1 + F \theta \right)}}{2\left(1- \theta B\right)^{2}}\left[\frac{1}{g^{2}}\left(E_{i}{}^{2} + B^{2}\right) + \left(\partial_{i}\phi\right)^{2}\right]
\label{statenergy}
\end{equation}
Note that the square root term in the first factor is taken positive by convention\footnote{ Note that this squareroot term can not be  written explicitly since depending on $\theta B < 1$ or $\theta B > 1$ one has to take $\sqrt{{\rm det}\left(1 + F\theta \right)} = \left(1 - \theta B\right)$ or $\left(\theta B - 1 \right)$.}. 
Evidently, from (\ref{statenergy}), minimum configuration of the static energy functional corresponds to the trivial solution where $E_{i}$, $B$ and $\partial_{i}\phi$ vanishes everywhere. These solutions also trivially satisfy the equations of motion. Clearly in the static limit there is no coupling between the matter and the gauge field. Hence there is {\it no} BPS soliton in the model. This result fills a gap in the existing literature regarding the possible B--P--S soliton solutions of the model and is consistent with the fact that there is no coupling of the matter and the gauge field in the corresponding commutative theory. Note however, that the solution space for the Bogomolony equations form a subspace of solutions of the equations of motion (\ref{eqmphi}, \ref{eqmA}) and our analysis does not rule out the existence of non-trivial non-B--P--S solutions of the equations of motion.

 
We have investigated a model involving the NC adjoint matter field coupled to a general NC $U(1)$ gauge field by an approach which complements the analysis of the model using the operator method. Our analysis was based on finding a commutative equivalent of the theory using S--W maps for the matter and gauge fields. A novel approach to analyze such theories in a closed form was presented. The equations of motion for the commutative equivalent of the proposed model were given. They are exact in the NC parameter $\theta$ without any restriction on the NC structure. Specializing the NC tensor by neglecting noncommutativity in the time-space direction, we have shown that the model does not have any nontrivial BPS soliton in the sector which has a smooth commutative limit. 
\section*{Acknowledgment}
AS would like to thank the Council for Scientific and Industrial Research (CSIR), Govt. of India, for financial support and the Director, S.N.Bose National Centre for Basic Sciences, for providing computer facilities. The authors also acknowledge the excellent atmosphere of IUCAA where part of the work has been done. They also thank the referee for useful comments.

\end{document}